\documentclass[useAMS,usenatbib]{mn2e}\voffset=-1cm
\usepackage{times,graphicx,aas_macros,psfig,rotating}
\DeclareGraphicsExtensions{.pdf,.png,.jpg,.ps}

\title[The Submm Properties of LBGs]{The SCUBA-2 Cosmology Legacy
  Survey: the submillimetre properties of
  Lyman break galaxies at \textit{z}=3--5}
\author[Coppin et al.]{\parbox[h]{\textwidth}{K.E.K.~Coppin$^{1}$\thanks{E-mail:
      k.coppin@herts.ac.uk}, J.E.~Geach$^{1}$, O.~Almaini$^{2}$,
   V.~Arumugam$^{3,4}$, J.S.~Dunlop$^{3}$, W.G.~Hartley$^{5}$, R.J.~Ivison$^{3,4}$,
C.J.~Simpson$^{6}$, D.J.B.~Smith$^{1}$, A.M.~Swinbank$^{7}$,
A.W.~Blain$^{8}$, N.~Bourne$^{3}$, M.~Bremer$^{9}$,  C.~Conselice$^{2}$, C.M.~Harrison$^{7}$,
    A.~Mortlock$^{3}$, S.C.~Chapman$^{10}$, L.J.M.~Davies$^{11}$,
    D.~Farrah$^{12}$, A. Gibb$^{13}$, T.~Jenness$^{14,15}$, A.~Karim$^{16}$,
    K.K.~Knudsen$^{17}$, E.~Ibar$^{18}$, M.J.~Micha{\l}owski$^{3}$,
     J.A.~Peacock$^{3}$, D.~Rigopoulou$^{19,20}$, E.I.~Robson$^{3,21}$, D.~Scott$^{13}$, 
    J.~Stevens$^{1}$, P.P.~van der Werf$^{22}$} \vspace*{6pt}\\ 
\noindent $^1$Centre for Astrophysics Research, Science \& Technology Research
Institute, University of Hertfordshire, College Lane, Hatfield AL10 9AB\\ 
$^2$The School of Physics and Astronomy, University of Nottingham, University Park, Nottingham NG7 2RD\\
$^3$SUPA, Institute for Astronomy, University of Edinburgh, Royal Observatory, Edinburgh EH9 3HJ\\
$^{4}$European Southern Observatory, Karl-Schwarzschild-Strasse 2, Garching,
D-85748, Germany\\
$^5$ETH Z\"{u}rich, Institut f\"{u}r Astronomie, HIT J 11.3,
Wolfgang-Pauli-Str. 27, CH-8093 Z\"{u}rich, Switzerland\\
$^6$Astrophysics Research Institute, Liverpool John Moores University, Twelve Quays House, Egerton Wharf, Birkenhead CH41 1LD\\
$^7$Institute for Computational Cosmology, Durham University, South Road, Durham DH1 3LE\\
$^8$Physics and Astronomy, University of Leicester, 1 University Road, Leicester LE1 7RH\\
$^9$Department of Physics, University of Bristol, H.H. Wills Physics
Laboratory, Tyndall Avenue, Bristol BS8 1TL\\
$^{10}$Dalhousie University, Department of Physics and Atmospheric Science, Coburg Road, Halifax, NS B3H 1A6, Canada\\
$^{11}$ICRAR, The University of Western Australia, 35 Stirling Highway, Crawley, WA 6009, Australia\\
$^{12}$Department of Physics, Virginia Tech, Blacksburg, VA 24061, USA\\
$^{13}$Department of Physics and Astronomy, University of British Columbia, 6224 Agricultural Road, Vancouver, BC V6T 1Z1, Canada\\
$^{14}$Department of Astronomy, Cornell University, Ithaca, NY 14853, USA\\
$^{15}$Joint Astronomy Centre, 660 N. A'Ohoku Place, University Park, Hilo, Hawaii 96720, USA\\
$^{16}$Argelander-Institut f\"{u}r Astronomie, Universit\"{a}t Bonn, Auf dem H\"{u}gel 71, D-53121 Bonn, Germany\\
$^{17}$Department of Earth and Space Sciences, Chalmers University of Technology, Onsala Space Observatory, SE-439 92 Onsala, Sweden\\
$^{18}$Instituto de Astrofisica, Facultad de Fisica, Pontificia Universidad Cat\'{o}lica de Chile, Casilla 306, Santiago 22, Chile\\
$^{19}$Department of Physics, Denys Wilkinson Building, Keble Road,
Oxford OX1 3RH\\
$^{20}$RAL Space, Science \& Technology Facilities Council, Rutherford
Appleton Laboratory, Didcot, OX11 0QX \\
$^{21}$UK Astronomy Technology Centre, Royal Observatory, Blackford Hill, Edinburgh EH9 3HJ\\
$^{22}$Sterrewacht Leiden, Universiteit Leiden, PO Box 9513, 2300 RA, Leiden, The Netherlands\\
}

\begin{document}

\date{}

\pagerange{\pageref{firstpage}--\pageref{lastpage}} \pubyear{2013}

\maketitle

\label{firstpage}

\begin{abstract} We present statistically significant detections at
850\,$\mu$m of the Lyman Break Galaxy (LBG) population at $z\approx3$, $4$ and
$5$ using data from the Submillimetre Common User Bolometer Array 2 (SCUBA--2)
Cosmology Legacy Survey (S2CLS) in the United Kingdom Infrared Deep Sky Survey
``Ultra Deep Survey'' (UKIDSS-UDS) field. We employ a stacking technique to
probe beneath the survey limit to measure the average 850\,$\mu$m flux density
of LBGs at $z\approx$3, 4, and 5 with typical ultraviolet luminosities of
$L_{\rm 1700}\approx10^{29}$\,erg\,s$^{-1}$\,Hz$^{-1}$. We measure 850$\mu$m
flux densities of (0.25$\pm$0.03), (0.41$\pm$0.06), and (0.88$\pm$0.23)\,mJy
respectively, and find that they contribute at most 20\,per cent to the cosmic
far-infrared background at 850\,$\mu$m. Fitting an appropriate range of
spectral energy distributions to the $z\sim3$, 4, and 5 LBG stacked
24--850\,$\mu$m fluxes, we derive infrared (IR) luminosities of
$L_{\mathrm{8-1000\mu m}}\approx3.2$, 5.5, and
11.0$\times$10$^{11}$\,L$_\odot$ (corresponding to star formation rates of
$\approx50$--200\,M$_\odot$\,yr$^{-1}$) respectively. We find that the
evolution in the IR luminosity density of LBGs is broadly consistent with
model predictions for the expected contribution of luminous IR galaxy (LIRG)
to ultraluminous IR galaxy (ULIRG) type systems at these epochs. We also see a
strong positive correlation between stellar mass and IR luminosity. Our data
are consistent with the main sequence of star formation showing little or no
evolution from $z=3$ to 5. We have also confirmed that, for a fixed mass, the
reddest LBGs (UV slope $\beta\rightarrow0$) are indeed redder due to dust
extinction, with SFR(IR)/SFR(UV) increasing by approximately an order of
magnitude over $-2<\beta<0$ such that SFR(IR)/SFR(UV)$\sim$20 for the reddest
LBGs. Furthermore, the most massive LBGs also tend to have higher
obscured-to-unobscured ratio, hinting at a variation in the obscuration
properties across the mass range. \end{abstract}


\section{Introduction}
Lyman Break Galaxies (LBGs) are currently the largest population of
star-forming galaxies known to be at high redshift, $z>3$, and as such have provided
valuable insights into the mass assembly of galaxies during the first few Gyr of the Universe. From a practical standpoint,
this has been due, in part, to the efficiency of the simple selection of LBGs
in broad-band colours that span the eponymous ``Lyman break'' as it is redshifted
through the optical bands at $z>3$ \citep{Steidel93}. This has made it possible to identify
large samples of LBGs, and from this we have learned that they are: massive
($M_\star\sim10^{9-11}$M$_\odot$; \citealt{Reddy06};
\citealt{Rigopoulou06}; \citealt{Verma07}; \citealt{Stark09}; \citealt{Magdis08,Magdis10a}), rapidly
star-forming (10s--100\,M$_\odot$\,yr$^{-1}$;
e.g.~\citealt{Shapley01,Shapley05}; \citealt{Magdis10b};
\citealt{Rigopoulou10}; \citealt{Chapman09}) and numerous ($\phi^{\ast}\sim
0.005$\,Mpc$^{-3}$; e.g.~\citealt{Reddy09}).   As such, they have been interpreted as a phase in the formation of
``typical'' galaxies (e.g.~\citealt{Somerville01}; \citealt{Baugh05}) and are the progenitors of a reasonable fraction of
massive ($L>L^\ast$) galaxies today.

Despite this progress, there are still a number of open questions surrounding
the nature and properties of typical LBGs. Some LBGs have dust-corrected star
formation rates (SFRs) of up to 100\,M$_\odot$\,yr$^{-1}$, which is
enough to form present-day elliptical galaxies, and indeed LBGs at $z>3$ are
attractive progenitors for the rather substantial population of passive red
galaxies already in place by $z\sim2$--3 (\citealt{Stark09}; \citealt{Kriek06};
\citealt{vanDokkum06}). Thus LBGs should contribute a significant portion of the
submm background at
850$\mu$m at high redshifts \citep{Adelberger00}. However,
determining their contribution to the submm background has been hampered by the large
uncertainties that go hand-in-hand with deriving dust-corrected UV
luminosities. The most reliable way of measuring their dust content and
contribution to the submm background is \textit{directly} through submillimetre (submm) observations, which will reveal any
dust-obscured star formation activity.

Several attempts have been made to detect the rest-frame far-IR emission in
LBGs, with mixed success. There are only a handful of individual submm
detections of LBGs, including Westphal-MMD11 \citep{Chapman00} and
Westphal-MM8 \citep{Chapman09}, targetted as part of a sample of LBGs with
high UV-derived SFRs, as well as a highly magnified gravitationally lensed LBG
at $z\sim3$ \citep{Baker01}. Two early SCUBA-based statistical studies of the
submm flux of LBGs with high UV-derived SFRs \citep{Chapman00} and of
``typical'' canonically-selected $z\sim3$ LBGs \citep{Webb03} yielded no
detection of the population, with an average below 1\,mJy. On the other hand,
\citet{Peacock00} detected a mean flux of (0.2$\pm$0.04)\,mJy for $0<z<6$
starburst galaxies with UV-SFRs of $>$2\,M$_\odot$\,yr$^{-1}$, with the mean
submm flux density increasing with UV-SFR. More recently with
\textit{Herschel}, several authors have revisited LBG stacking. \citet{Lee12}
find a marginal signal ($\sim3$--4\,$\sigma$ level) in a single-band
(500\,$\mu$m) for a UV-bright subset of $z\sim4$ LBGs, and
\citet{Rigopoulou10} and \citet{Magdis10b} have reported statistical
detections of mid-IR-selected LBGs with \textit{Herschel}, but this latter
sample is clearly biased to the IR-brightest most massive and/or dustiest
subset of LBGs (\citealt{Huang05}; \citealt{Rigopoulou10}; see also
\citealt{Oteo13}). As the available submm maps continue to increase in
size and depth (e.g.~\citealt{Weiss09}), stacking is now able to
sometimes yield successful statistical detections of specially chosen subsets of
LBGs at $z\sim3$ (split by stellar mass and UV-estimated SFRs;
\citealt{Davies13}), which has helped to reliably constrain the far-IR
luminosities, obscured SFRs, dust masses and dust temperatures in the most
massive and UV-luminous $z\sim3$ LBGs. There have also been attempts to detect
the dust emission in a handful of LBGs at $z\sim5$ (e.g.~\citealt{Stanway10};
\citealt{Davies12}), although no detections were made, making it difficult to
say anything conclusive. Thus, the far-IR emission from \textit{typical} LBGs
(i.e.\ those selected by the canonical UV/optical broadband colour criteria),
hence their dust content and contribution to the submm background, as well as tracing
their IR luminosity density as a function of redshift all remain open
questions.

Here we present a statistical (stacking) analysis of LBGs at $z\sim3$, 4 and 5
in the United Kingdom Infrared Deep Sky Survey ``Ultra Deep Survey''
(UKIDSS-UDS; \citealt{Lawrence07}) field. The UDS has been mapped with the
\textit{Herschel} Spectral and Photometric Imaging Receiver (SPIRE) at 250,
350, and 500\,$\mu$m and at 850\,$\mu$m as part of the SCUBA--2
\citep{Holland13} Cosmology Legacy Survey (S2CLS; e.g.~\citealt{Geach13}). The
850$\mu$m map is the largest submm map yet obtained at this depth
($1\sigma\approx2$\,mJy over 1\,deg$^2$, see \S2.1). The availability of deep
optical/near-infrared (OIR) imaging in this field, combined with the new
wide-area deep 850\,$\mu$m map makes this an ideal resource to study the submm
properties of LBGs, without biasing the analysis to the most massive and/or
mid-IR or UV-brightest LBGs (e.g.~\citealt{Rigopoulou10}; \citealt{Davies13}).
The goal of this work is to measure the average submm flux density (which is a
direct probe of obscured star formation) of LBGs selected at $z=3$, $4$ and
$5$ in order to estimate the bolometric luminosity of typical massive
galaxies at these epochs and to assess the LBG contribution to the IR luminosity
density to compare with models.

This paper is organized as follows.  The data and LBG sample selections
are described in Section~2.  In Section~3 we present the main analysis
and results of the LBG stacking in submm maps, including the LBG
contribution to the submm background, LBG SEDs, and the evolution of
the IR luminosity density of LBGs.  In Section~4 we discuss the
implications of the results and, finally, our conclusions are given in Section~5. 
Throughout the paper we assume cosmological parameters of
$\Omega_\Lambda=0.73$, $\Omega_\mathrm{m}=0.27$, and
$H_\mathrm{0}=71$\,km\,s$^{-1}$\,Mpc$^{-1}$ \citep{Spergel03}.

\section{The Data}

\subsection{Optical and near-infrared imaging}

Our samples are based on the deep $K$-band image from the United Kingdom
Infrared Telescope (UKIRT) Infrared Deep Sky Survey (UKIDSS;
\citealt{Lawrence07}) Ultra Deep Survey
(UDS\footnote{http://www.nottingham.ac.uk/astronomy/UDS/}) data release 8
(DR8; Almaini et al., in preparation) and co-incident multi-wavelength data.
The parent catalogue was extracted using {\sc sextractor} \citep{Bertin96} on
the deep $K$-band image ($K_\mathrm{AB}\leq24.6$) and after merging the source
lists from two independent extractions. The extraction parameters were
designed to recover both point-like and extended low-surface-brightness
objects (see \citealt{Hartley13} for further details). In addition to the
three $JHK$ UKIDSS bands, the UDS has been mapped by the Canada-France-Hawaii
Telescope (CFHT) Megacam $u'$ band ($u'_\mathrm{AB}\leq26.75$; Foucaud et al.,
in preparation), by Subaru Suprime-cam in the optical ($B = 27.6$, $V=27.2$,
$R=27.0$, $i'=27.0$ and $z'=26.0$, 5-$\sigma$ AB mags; \citealt{Furusawa08})
and by the \textit{Spitzer} InfraRed Array Camera (IRAC; \citealt{Fazio04}) in channels 1
(3.6\,$\mu$m; 5-$\sigma$ depth of 24.2 AB mags), and 2 (4.5\,$\mu$m;
5-$\sigma$ depth of 24.0 AB mags) via the UDS \textit{Spitzer} Legacy Program
(SpUDS; PI:Dunlop). The overlap of all these data sets (after masking bad
regions) is 0.62\,deg$^{2}$. In addition, available X-ray \citep{Ueda08} and
radio \citep{Simpson06} data were employed to remove bright AGN. The matched
multi-wavelength photometry was extracted using 3\,arcsec apertures centred on
the positions of the $K$-band catalogue sources (see \citealt{Simpson12} for
full details). Point spread function (PSF) photometry corrections were
required and applied to three of the bands (the CFHT $u'$ band and the two
IRAC channels) in order to obtain correct colours (see \citealt{Hartley13} for
details). We use this final matched multi-wavelength catalogue to perform our
LBG selection.

\subsubsection{Lyman Break Galaxy selection}

The canonical LBG selection utilizes the {\it UGR}, or {\it BVR}, colour space,
with ``drop-outs'' in the bluest band efficiently isolating galaxies at
$z\approx3$ \citep{Steidel96}. Since the selection requires detection
in {\it G} and {\it R}, as the Lyman break is redshifted through the {\it G} or {\it V} band, the selection
function falls off with redshift. The technique can be applied to higher redshift by simply
moving the entire colour space to longer wavelengths. \citet{Ouchi04} show how
a similar selection in {\it BRi}, isolates LBGs at $z\approx4$, and {\it Viz}, or {\it Riz}, extends the selection to $z\approx5$.

We adopt the following colour selections for LBGs at $z\approx3$ (eq.~1),
$z\approx4$ (eq.~2) and $z\approx5$ (eqs.~3 and 4):

\begin{center}
\begin{equation}
\begin{array}{ll}
R<27, & (U -V)>1.2, \\
-1.0<(V -R)<0.6, &
(U-V)>3.8 (V -R) + 1.2;\\	
\end{array}
\end{equation}
\end{center}

\begin{center}
\begin{equation}
\begin{array}{ll}
I<27, & (B -R)>1.2, \\
(R -I)<0.7, & 
(B -R)>1.6(R -I) + 1.9;\\	
\end{array}
\end{equation}
\end{center}

\begin{center}
\begin{equation}
\begin{array}{ll}
Z<26, & (V-I)>1.2, \\
(I -Z)<0.7, & 
(V -I)>1.8(I-Z) + 2.3;\\	
\end{array}
\end{equation}
\end{center}

\begin{center}
\begin{equation}
\begin{array}{ll}
Z<26, & (R -I)>1.2, \\
(I -Z)<0.7, & 
(R -I)>(I -Z) + 1.0.\\	
\end{array}
\end{equation}
\end{center}

\noindent For the $z=5$ selection, we require selection using {\it either} eq.~3 or eq.~4,
slightly improving our yield (see Ouchi et al.\ 2004), and for all selections
we also require that the source has not been identified as a possible star.

Eleven-band photometric ({\it UBVRIzJHK}[3.6][4.5]) redshifts have been estimated
for all of the galaxies in the DR8 parent sample. The overview of these redshifts and how they
are calculated is discussed in more detail in \citet{Hartley13} and
\citet{Mortlock13}.  In summary, the photometric redshifts 
are computed using the {\sc eazy} template fitting code, adopting the six default
{\sc eazy} spectral energy distribution (SED) templates
\citep{Brammer08}. In addition to this we further include a template
which is the bluest {\sc eazy} template, but with a small amount of
Small Magellanic Cloud (SMC) like extinction added. Photometric
redshifts are then computed from a maximum likelihood analysis.  To
test how accurate these photometric redshifts are, we compare the
values we calculate to spectroscopic redshifts that are available in
the UDS. A large fraction of these are from the UDSz, an European
Southern Observatory (ESO) large
spectroscopic survey (ID:180.A-0776; Almaini et al., in preparation) and also from previous published
values (see \citealt{Simpson12} and references therein).  We have in
total about 1500 secure spectroscopic redshifts in the UDSz, and
around 4000 redshifts from other sources, although this reduces to 2146
after the removal of active galactic nuclei (AGN).  Excluding
catastrophic outliers ($|z_{\rm phot} - z_{\rm spec}|/(1 + z_{\rm
  spec})>$\,0.15), we find an average $|z_{\rm phot} - z_{\rm spec}|/(1 + z_{\rm spec})=0.031$ \citep{Hartley13}.   

\begin{figure} \includegraphics[height=\linewidth,angle=-90]{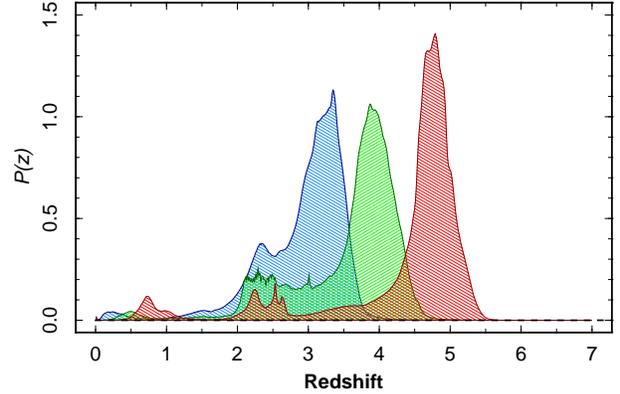} \caption{Stacked
photometric redshift distributions from {\sc eazy} (see text) of the sources
selected in our three LBG samples (all normalised to a common integral),
clearly showing the three distinct peaks in redshift space, centred at
$z\sim3$, 4, and 5.} \label{photoz} \end{figure}

To further improve the LBG colour selections described above, we enforce a
photometric redshift $z_{\rm phot}\geq2$ selection to eliminate any low-{\it
z} interlopers (where $z_{\rm phot}$ is the solution with the minimum
$\chi^{2}$ from the {\sc eazy} fitting). The photo-{\it z} fitting procedure evaluates a probability
density distribution, $P(z)$, for each galaxy, with the peak of that
distribution representing the best estimate for the redshift. Apart from the
$z_{\rm phot}\geq2$ criterion, we do not use the photometric redshifts
further. However, we can make use of $P(z)$ to investigate the efficacy of the
colour selections described above. In Fig.~\ref{photoz} we show the
(normalised) sum of the individual $P(z)$ for the galaxies in the selections
defined in equations 1--4. This clearly demonstrate the effectiveness of the
colour selections, and also provides us with an estimate of the widths of the
redshift distributions, reflecting both intrinsic spread (due to the
``window'' where the redshifted Lyman break can be identified in each case),
and uncertainty in the photometric redshift estimates. The resulting LBG
sample properties are summarized in Table~1.

An important point to note is that our LBG selection is predicated on a
$K<24.6$ selected sample, that is the basis of the UKIDSS-UDS catalogue
described above. As a result, we will not include, for example, $z\approx3$
LBGs with $R<27$ and $K>24.6$, but these will be at the low stellar mass end
of the distribution ($M_\star<10^9M_\odot$, see \S~\ref{stellarmass}). As we explore in
\S~\ref{stellarmass}, the average submm flux of LBGs is expected to be a function of
mass, and so we treat the $K>24.6$ limit as a proxy mass limit for the
galaxies in the sample. Note also that the optical limits in equations
1--4 result in a slight bias in UV luminosity selection with
redshift. At $z=3$ we are sensitive to galaxies more luminous than
$M_{\rm 1700}\sim-19$\,mag, but only $M_{\rm 1700}\sim-20$\,mag at
$z\sim5$. Therefore, when we examine trends with UV luminosity, we
focus on the $z\sim3$ sample.

\subsection{SCUBA--2 observations and map making}\label{scuba}

As part of the S2CLS campaign, observations of UKIDSS-UDS were conducted in Band 2/3 weather
($0.05<\tau_{\rm 225\,GHz}\leq0.1$) between October
2011 and February 2013. The mapping centre of the UDS field is
$\alpha=$\,2$^{\rm h}$\ 18$^{\rm m}$\ 00$^{\rm s}$,
$\delta=$\,$-$05$^{\circ}$\ 05${'}$\ 00${''}$, and a 3300\,arcsec diameter PONG
mapping pattern was used, resulting in a circular map that reaches a uniform
depth over an area of $\sim$700 arcmin$^2$. The total mapping time was
130\,hr, split into 195 scans of 40 minutes each. Individual scans are reduced
using the dynamic iterative map-maker ({\it makemap}) of the {\sc smurf}
package \citep{Chapin13}. \citet{Geach13} presented the reduction procedure
for CLS data, but in summary, we apply the following steps.

\begin{enumerate}
\item Raw data are flat-fielded using ramps bracketing every science
observation, and the data are scaled to units of pW.

\item The signal each bolometer records is then assumed to be a linear
combination of: (a) a common mode signal dominated by atmospheric water and
ambient thermal emission; (b) the astronomical signal (attenuated by atmospheric
extinction, see \citealt{Dempsey13}); and finally (c) a noise term
(including $1/f$-type correlations), taken to
be the combination of any additional signal not accounted for by (a)
and (b). The
dynamic iterative map maker attempts to solve for these model components,
refining the model until convergence is met, or an acceptable tolerance has
been reached.

The reduction also includes the usual filtering steps of spike removal
($>$10$\sigma$ deviations in a moving boxcar) and DC step corrections.
Throughout the iterative map-making process, bad bolometers (those
significantly deviating from the model) are flagged and do not contribute to
the final map.

\item Time streams are finally re-gridded into a map with 4\,arcsec pixels,
with each bolometer weighted according to its time-domain variance (which is
also used to estimate the $\chi^2$ tolerance in the fit). \end{enumerate}

Maps made from individual scans are co-added in an optimal, noise-weighted
manner, using the {\sc mosaic\_jcmt\_image} recipe in the {\sc Picard}
environment \citep{Jenness08}. Finally, to improve the detectability of faint
point sources, we apply a beam-matched filter to improve their detectability
using the {\sc Picard} recipe {\sc scuba2\_matched\_filter}. The average
exposure time over the nominal 3300\,arcsec mapping region in the co-added map
is approximately 1.5\,ksec per 4$''$ pixel, and the r.m.s. noise value is
1.9\,mJy. Due to the scanning strategy, there are data beyond the 3300\,arcsec
perimeter, but because these receive less total exposure, the noise increases
accordingly. The LBG catalogues cover a 0.6\,deg$^{2}$ area within the central
uniform noise region of the SCUBA-2 map.

An important component of our stacking analysis is to create a map with bright
significantly-detected point sources near to an LBG (but not near enough to be
associated) removed. For this, we run a peak-finding algorithm on the
signal-to-noise ratio map down to a level of 3.5\,$\sigma$, and this was done
for each LBG-redshift sub-sample separately. Each time a peak is found within
8\,arcsec (about half a SCUBA-2 beam), the flux of the contributing source is
removed by subtracting a model point response function (PRF) at that position,
scaled to the flux of the source. The peak-normalised model PRF is generated
from the data by stacking the map at the positions of all $>$10$\sigma$ peaks.
This residual map still contains astronomical signal, but with deemed
LBG-unassociated submm $>$3.5\,$\sigma$ sources removed, so that we do not
inadvertently bias our stacking results high by including flux from nearby
unrelated positive beams. This procedure is virtually identical to the
approach taken in the \citet{Webb03} SCUBA LBG stacking analysis paper. The
remaining 36 submm sources\footnote{We find 26, 9, and 1 candidate LBG-SMG
counterparts for our LBG redshift sub-samples at $z\sim3$, 4, and 5,
respectively.} may not be genuine LBG-SCUBA-2 associations (and in fact the
calculated Poisson probabilities that they are chance coincidences are not
particularly low), but because they \textit{could} be genuine, we must not
remove these sources from the map. These candidate LBG-SCUBA-2 associations
will be evaluated using a rigorous multi-wavelength approach in Coppin et al.\
(in preparation).

\section{Analysis \& Results}

\subsection{Statistical results}\label{stack}

\begin{figure}
	\includegraphics[width=\linewidth]{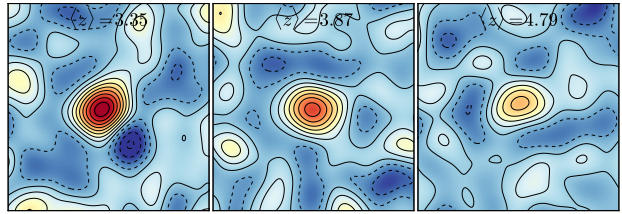}
	\caption{80\,arcsec $\times$ 80\,arcsec thumbnail images of
          signal-to-noise ratio of the stacked 850\,$\mu$m flux in the
          SCUBA-2 map, centred on the LBG positions in the $z\sim3$, 4, and 5 sub-samples (the mean redshift of each LBG sub-sample is indicated
          on each image).  The solid contours represent
          0, 1, 2, 3-$\sigma$,..., and the dashed lines are the
          corresponding negative contours.}
\label{thumbnail}
\end{figure}

With the LBG samples defined, we measure the submm flux at each LBG position
in the 850\,$\mu$m beam-convolved, $\geq$3.5$\sigma$ source subtracted flux
map, weighted by the noise at the corresponding positions in the noise map.
For visualisation, we simultaneously stack 80$\times$80\,arcsec cut-outs,
again applying a simple inverse variance weighting scheme, obtaining the
weights from the noise maps (note that measuring the stacked flux directly
from these images necessarily gives the same result). The stacked thumbnail
images for each sample are shown in Fig.~\ref{thumbnail}, indicating the
detection of significant 850$\mu$m emission from LBGs at $z\approx3$, $4$, and
$5$ (see Table~1). We repeat the exercise above on the original flux map
(i.e.\ including all $>$3.5\,$\sigma$ sources which had been removed) and list
these results in Table~1. These stacked fluxes are statistically
indistinguishable from the stacked fluxes calculated previously on the
source-subtracted maps. As an additional check, we compare the radially
averaged profiles of the stacks with the shape of the beam (empirically
derived by stacking many high significance peaks in the detection image). The
stacks are indistinguishable from the beam, and this indicates that if there
is any significant contribution to the observed signals from large scale
structure correlated with the LBGs, it is on scales below $\sim$100\,kpc.

To verify the significance of our results, we have stacked at the same number
of uniform random positions within the same map area and repeated this 10,000
times for each LBG redshift slice subset. We do not, however, remove any
sources from the map; this should provide a conservative estimate of the
likelihood of measuring the same (or greater) level of flux in the stack as
observed for the LBGs simply by chance. In these random realisations, we never
find a result as statistically significant as ours in the $z\sim3$ and 4
samples (the most significant stacked result we find at random in the $z\sim$3
and 4 samples are at the 3.8 and 3.4\,$\sigma$ levels, respectively, occurring
in each $<$0.02 per cent of the time), and less than 0.02 per cent of the time
in the $z\sim5$ sample. Thus, in light of these simulations our statistical
detections appear to be robust.

\begin{table*}

\centering

\caption{Stacked mid-IR-to-submm photometry of Lyman Break Galaxies}

\hspace{-0.2in}

\begin{tabular}{cccrrrrrl}
\hline

\multicolumn{1}{c}{$\langle z_\mathrm{phot}\rangle$} & \multicolumn{1}{c}{$N$}  &  \multicolumn{1}{c}{850\,$\mu$m$^{a}$} & \multicolumn{1}{c}{850\,$\mu$m$^{b}$} & \multicolumn{1}{c}{250\,$\mu$m} & \multicolumn{1}{c}{350\,$\mu$m} &\multicolumn{1}{c}{500\,$\mu$m} & \multicolumn{1}{c}{24\,$\mu$m}\\
 \multicolumn{1}{c}{} &
\multicolumn{1}{c}{}  & \multicolumn{1}{c}{(mJy)} &
\multicolumn{1}{c}{(mJy)} &
\multicolumn{1}{c}{(mJy)} &
\multicolumn{1}{c}{(mJy)}&
\multicolumn{1}{c}{(mJy)}& \multicolumn{1}{c}{($\mu$Jy)}\\
\hline
  3.35 & 4201  & 0.249$\pm$0.029
(8.5$\sigma$) & 0.228$\pm$0.029 (7.8$\sigma$) & 0.751$\pm$0.077 (9.8$\sigma$)&
0.938$\pm$0.109 (8.6$\sigma$) & 0.634$\pm$0.143 (4.4$\sigma$) &
9.5$\pm$0.9 (10.5$\sigma$)\\

3.87 & 869 &    0.411$\pm$0.064
(6.4$\sigma$) & 0.388$\pm$0.064 (6.0$\sigma$) & 0.783$\pm$0.177 (4.4$\sigma$)&
1.116$\pm$0.163 (6.8$\sigma$) & 1.024$\pm$0.164 (6.2$\sigma$) &
2.1$\pm$0.5 (4.2$\sigma$)\\

4.79 & 68   & 0.875$\pm$0.229
(3.8$\sigma$) & 0.811$\pm$0.229 (3.5$\sigma$) & 2.023$\pm$0.798 (2.5$\sigma$)&
2.337$\pm$0.916 (2.6$\sigma$) & 2.022$\pm$0.686 (2.9$\sigma$) &
4.0$\pm$2.0 (2.0$\sigma$)\\
\hline
\end{tabular}
$^{a}$ Submm sources $\geq$3.5\,$\sigma$ and further than 8\,arcsec
away from an LBG have been removed. We adopt this column for the SED fitting.\\
$^{b}$ No submm sources have been removed.

\end{table*}

\begin{table*}

\centering

\caption{Derived average IR and 1700\AA\ luminosities and corresponding
  obscured (IR) and unobscured (UV) SFRs of $z\sim3$--5 Lyman
  Break Galaxies}

\hspace{-0.2in}

\begin{tabular}{crrcc}
\hline

\multicolumn{1}{c}{$\langle z_\mathrm{phot}\rangle$} &
\multicolumn{1}{c}{$L_\mathrm{IR}$} &
\multicolumn{1}{c}{SFR$_\mathrm{IR}$} & \multicolumn{1}{c}{$L_\mathrm{1700}$} & \multicolumn{1}{c}{SFR$_\mathrm{UV}$} \\
\multicolumn{1}{c}{} & \multicolumn{1}{c}{(10$^{11}$\,L$_\odot$)} & \multicolumn{1}{c}{(M$_\odot$\,yr$^{-1}$)} & \multicolumn{1}{c}{(10$^{29}$\,erg\,s$^{-1}$\,Hz$^{-1}$)} &
\multicolumn{1}{c}{(M$_\odot$\,yr$^{-1}$)}\\
\hline
3.35 & 3.2$^{+0.8}_{-0.6}$ & 55$^{+14}_{-10}$ & 0.9 & 18\\
3.87 & 5.5$^{+0.3}_{-0.4}$ & 93$^{+5}_{-7}$ & 1.7 & 33\\
4.79 & 11.0$^{+4.2}_{-5.9}$ & 186$^{+71}_{-101}$ & 1.7 & 33\\
\hline
\end{tabular}
\end{table*}

\subsubsection{Assessing the level of bias in our statistical results}

Our simple methodology implicity assumes that the galaxy population we have
stacked is uncorrelated (i.e.\ the LBGs are not clustered). However, like most
populations, LBGs are known to be clustered
($r_\mathrm{0}\simeq4\,h^{-1}$\,Mpc; \citealt{Adelberger05}). We investigate
the level of bias due to LBG clustering in our stacking results by following
the formalism of \citet{Viero13} for unbiased stacking of galaxy catalogues on
\textit{Herschel} maps (see also \citealt{Kurczynski10} and
\citealt{Roseboom10}), which takes into account the presence of multiple
sources in the same beam. In brief, we create a ``hits'' map ($H$) of the same
size as the SCUBA-2 map for a given LBG catalogue, where each pixel in the map
contains the integer number of sources that fall into it. The hits map is then
convolved with the \textit{actual} PRF of the instrument (note that a
different result, yielding a non-negligible positive flux bias, occurs if the
beam is wrongly assumed to be a simplistic Gaussian). The convolved hits map
is then regressed against the 850\,$\mu$m flux map ($S_{850}$) to find the
minimal flux residual $|\langle S \rangle H-S_{850}|$, where $\langle S
\rangle$ is the average flux density of the LBG population. The method safely
assumes that galaxies in different redshift slices are uncorrelated and we
thus treat each of our LBG lists independently. The fitting routine yields
minima at $S_\mathrm{850}$=(0.21$\pm$0.04), (0.39$\pm$0.07), and
(0.86$\pm$0.26)\,mJy for $z\sim3$, 4, and 5, respectively (with the error bars
calculated from Monte Carlo simulations). These results are indistinguishable
from the results in columns 3 and 4 of Table~1. It is reassuring that our statistical
detections do not appear to be dominated by an upwards bias in flux introduced
by potentially ``double counting'' flux for multiple sources falling in the
same beam.

There is also the legitimate worry that LBGs are also correlated with some
fainter (as yet undetected, and potentially numerous) population that
contributes to the observed submm flux density, thus boosting our statistical
detections (e.g.~\citealt{Chary10}, \citealt{Serjeant10}, and
\citealt{Kurczynski10}). However, for the purposes of our analysis we would
say this was flux associated with the LBG. Whether or not the emission is
always physically within the LBG is beyond the scope of our current study.

\subsection{The LBG contribution to the submm background at 850$\mu$m}

The contribution to the submm background from LBGs is still poorly
constrained, however our data can finally address this question, since we now
have a robust detection of the average 850\,$\mu$m flux density of LBGs at
three epochs, at least those with ultraviolet luminosities of
$L_{1700}\approx10^{29}$\,erg\,s$^{-1}$\,Hz$^{-1}$, characteristic of galaxies
in our sample (Table\ 2). Using our average stacked 850\,$\mu$m flux
densities, we estimate surface brightness densities of 1700, 600, and
100\,mJy\,deg$^{-2}$ of LBGs at $z\sim3$, 4, and 5, respectively. By
comparison, the total background at 850\,$\mu$m inferred from
\textit{COBE}-FIRAS is 3.1--$4.4\times10^{4}\,\mathrm{mJy}\,\mathrm{deg}^{-2}$
(\citealt{Puget}; \citealt{Fixsen}; \citealt{Lagache_rev};
\citealt{Hauser01}). Summing these separate surface brightness densities
together, we find that the LBGs in our $z\sim3$, 4 and 5 samples comprise
around 6--8\,per cent of the submm background at 850\,$\mu$m (where the range
of values simply reflects the uncertainty in the \textit{COBE}-FIRAS result).
However, the true contribution from LBG-like galaxies to the submm background
will come from a wider range in redshift, not just from the rather narrow
redshift slices we have sampled (see Fig.~\ref{photoz}), and from sources
falling out of our samples due to incompleteness. To determine the total
(corrected) contribution from LBGs over $3<z<5$, we assume that LBGs have a
constant comoving number density (which is a reasonable assumption since the
bright end of the luminosity function for LBGs shows little evolution over
$3<z<5$; \citealt{Reddy09}; Bouwens{07}; \citealt{McLure09}). Starting with
the $z\sim3$ sample (which is our most complete sub-sample of LBGs) we can
integrate over the comoving volume element for this redshift range, scaling
the LBG submm background contribution accordingly. We find that the total
contribution to the submm background from LBGs over the redshift range $3<z<5$
is likely to be closer to 14--20\,per cent. This result is consistent with
\citet{Webb03}, who estimated an upper limit to the contribution to the submm
background from $1<z<5$ of less than 20\,per cent.

\subsection{The bolometric luminosities and total SFRs of LBGs}\label{seds}

\begin{figure*}
\includegraphics[width=0.35\textwidth,angle=-90]{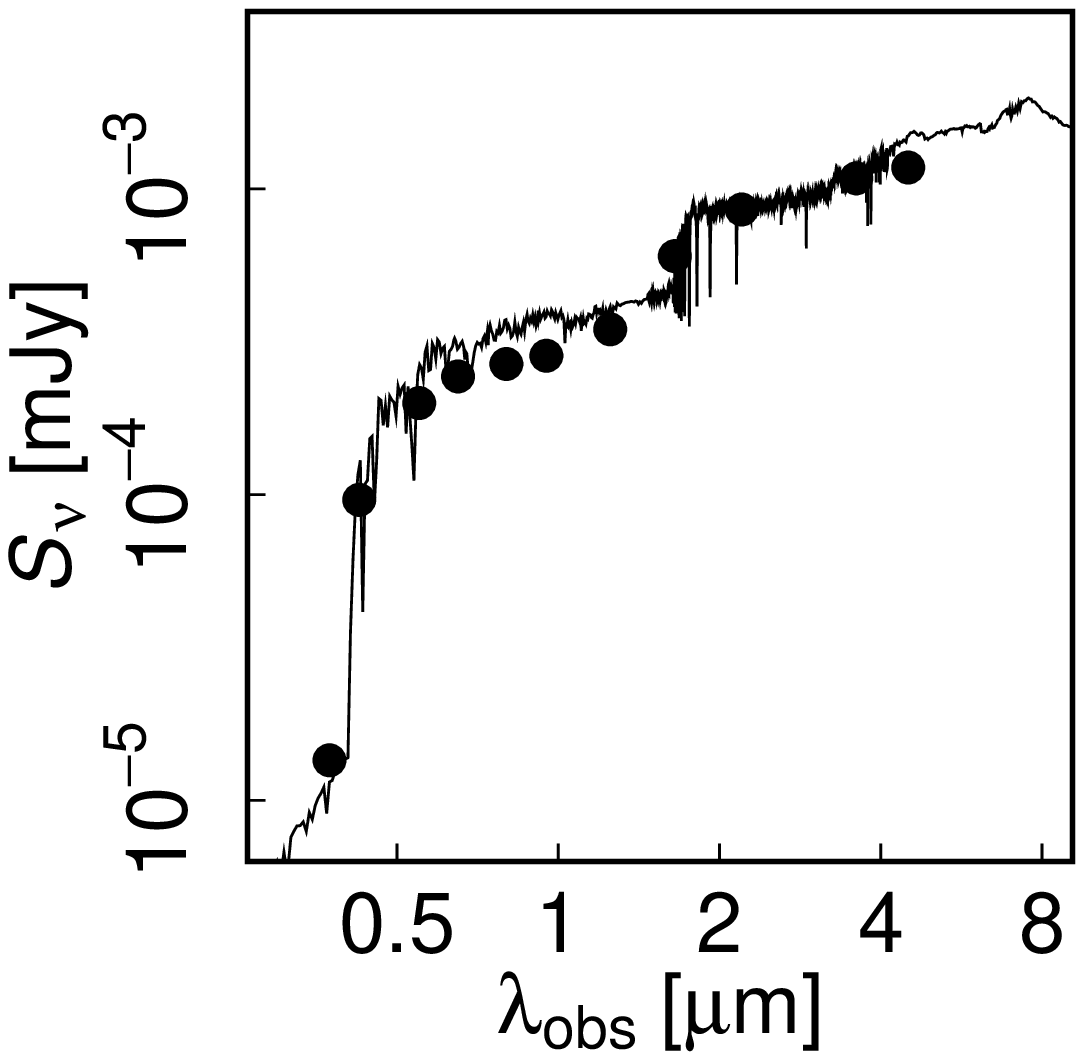}\includegraphics[width=0.35\textwidth,angle=-90]{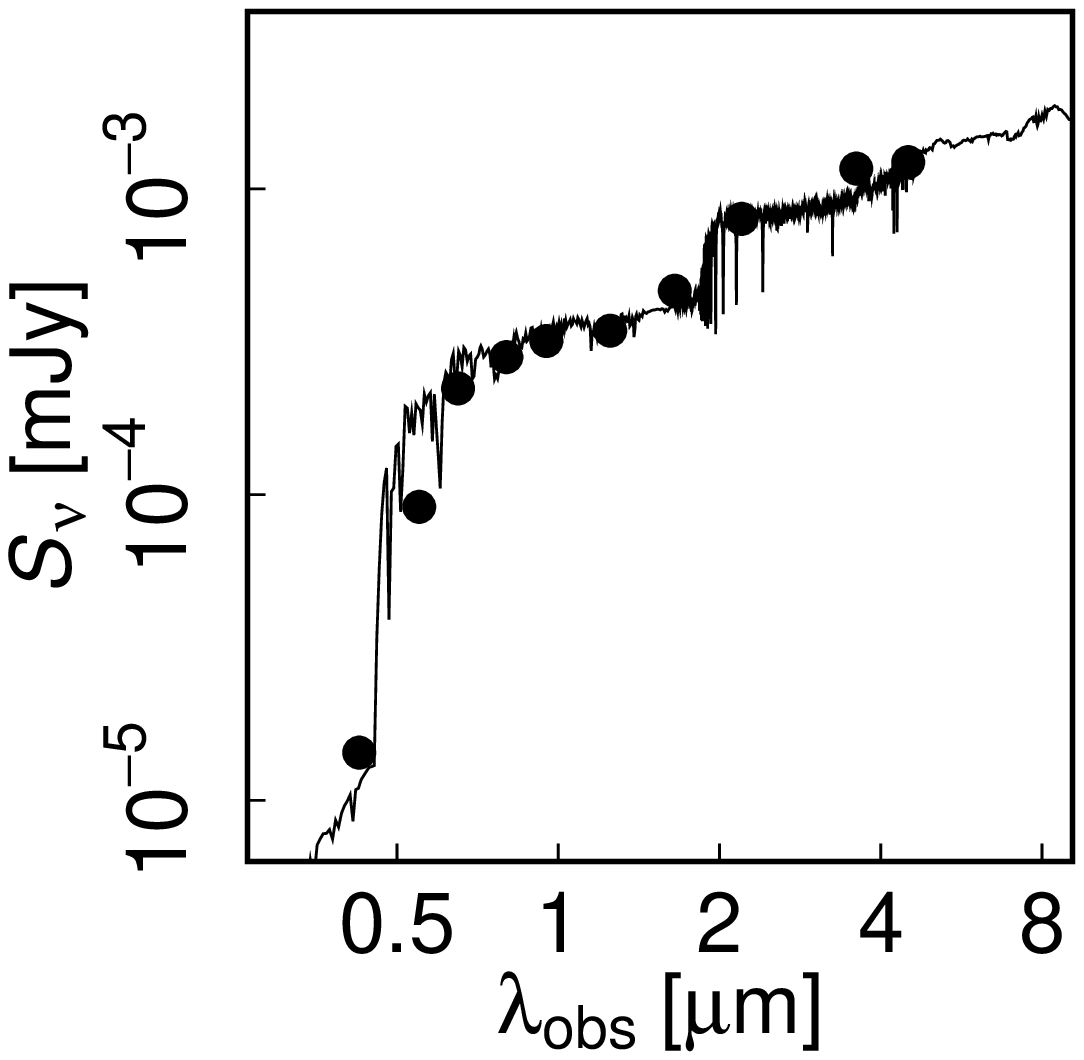}\includegraphics[width=0.35\textwidth,angle=-90]{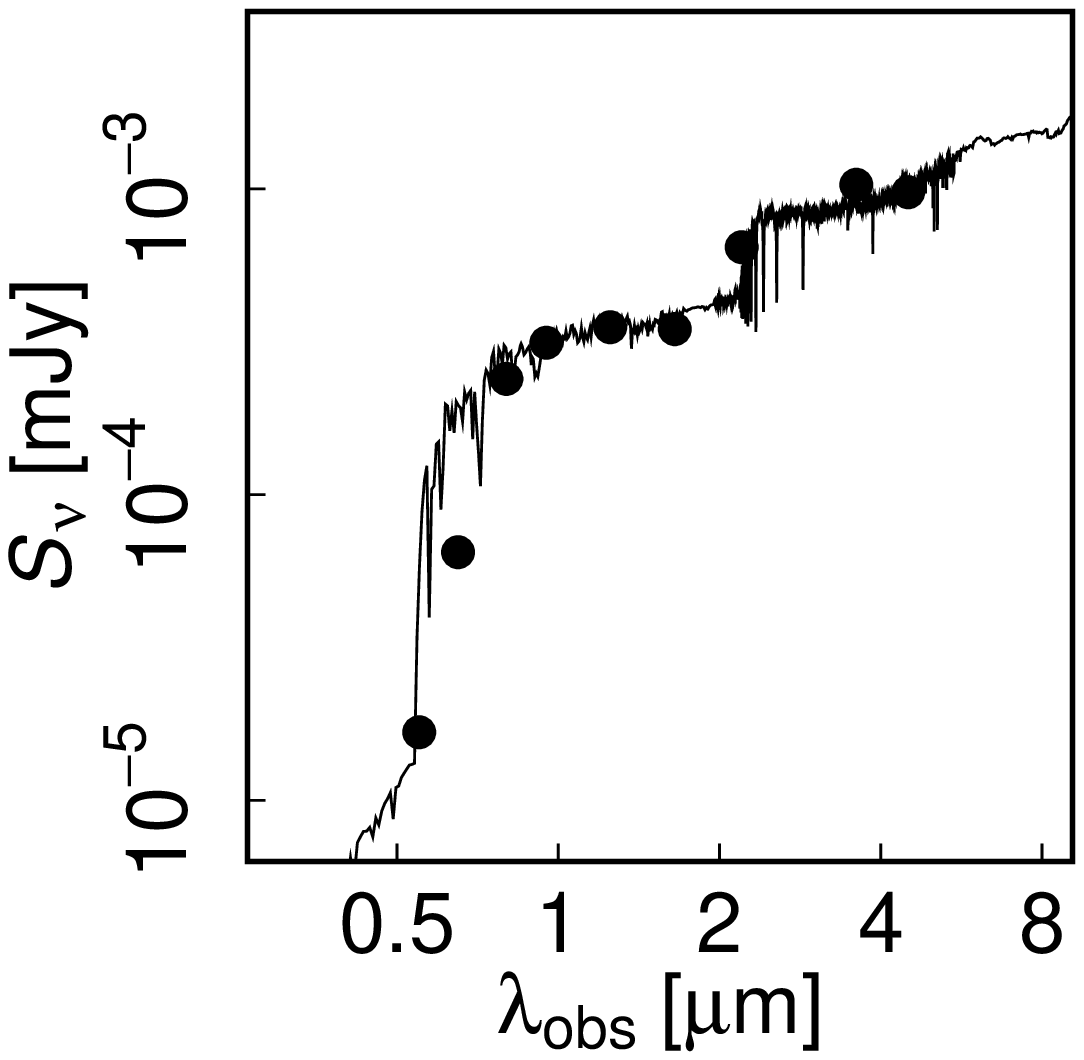}
\caption{Best-fitting \citet{Bruzual07} model for $z=3$ LBGs selected
  at 24$\mu$m from \citet{Magdis10b}, 
normalised to the observed $K$-band flux of each LBG sample at $z=3$, 4, and 5.  This
simple SED scaling provides an overall excellent description
of the observed average photometry, with only a slight
discrepancy in the UV emission at $z=4$ and $z=5$, most likely due to
increasing Ly$\alpha$ forest absorption at $z>3$ (see text).}
\label{optseds}
\end{figure*}

\begin{figure*}
\includegraphics[height=\textwidth,angle=90]{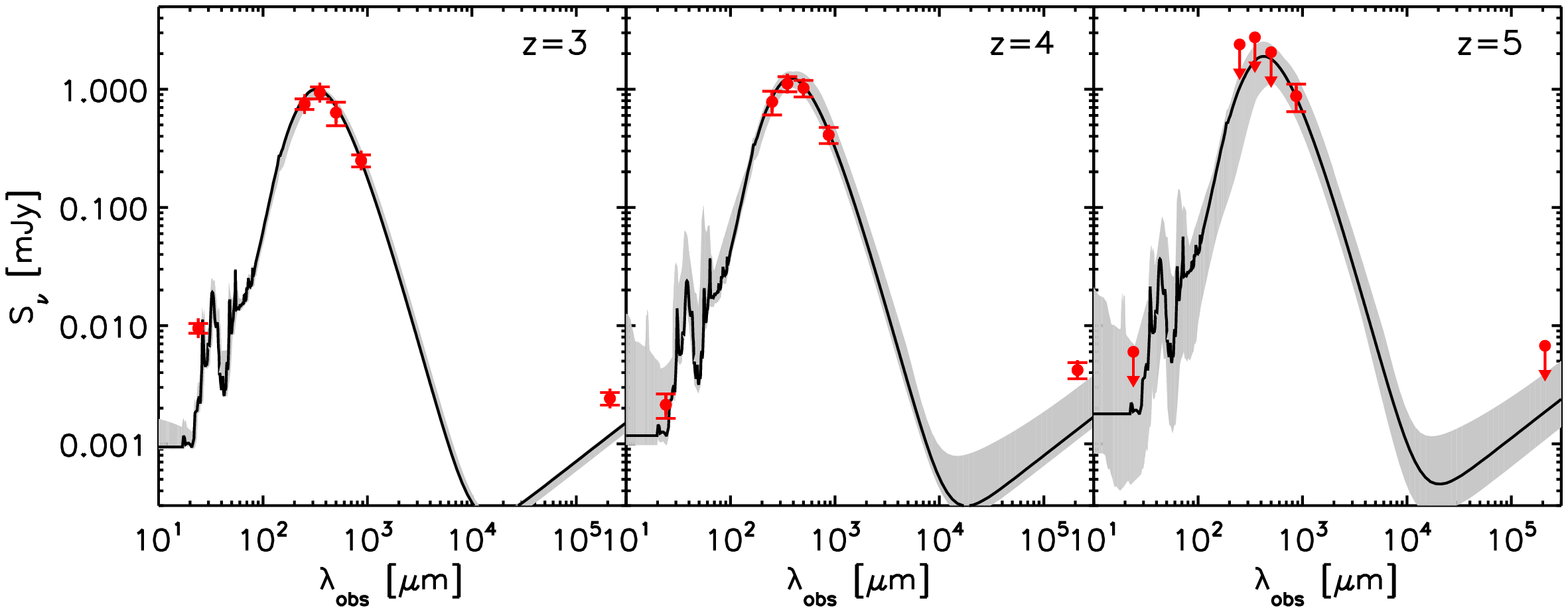}
\caption{Best-fitting far-IR SEDs of the LBG
  samples at $z=3$, 4, and 5, with the shaded regions showing the
  range of acceptable solutions
  (see text).  We have used the stacked fluxes (or 3\,$\sigma$ upper limits, where
  appropriate) at 24, 250, 350, 500 and
  850\,$\mu$m in the fitting. Also overplotted are the median stacked
1.4\,GHz radio fluxes, which appear to lie mildly (1--$3\,\sigma$) higher than
the best-fitting far-IR constrained SEDs would predict for the
$z\sim3$ and 4 samples, hinting
that either the radio emission in the template is not representative of $z\sim3$--4 star-forming
galaxies and/or that there is some AGN
contribution to the radio fluxes of typical LBGs (see text).}
\label{far-ir-seds}
\end{figure*}

In order to put LBGs in context with other IR-luminous galaxy populations we
must determine some intrinsic properties, such as their IR luminosities and
obscured and unobscured SFRs. Our approach is to make use of the full
(rest-frame UV to submm) average spectral energy distributions (SEDs) of LBGs
at each epoch, where the photometry are the average for the sample.

For the average rest-frame UV--near-IR SED of our LBGs, we find that the best
fitting \citet{Bruzual07} model for $z=3$ LBGs selected at 24\,$\mu$m-selected
from \citet{Magdis10b} provides a reasonable model of the stellar emission in
our {\it non}-IR selected sample. In Fig.~\ref{optseds} we show the average
photometry of the three LBG samples, with the \citet{Magdis10b} SED redshifted and
scaled to the observed $K$-band flux. The template SED provides an excellent
description of the observed average photometry, with only a slight discrepancy
in the UV emission at $z=4$ and $z=5$, most likely due to increasing
Ly$\alpha$ forest absorption at $z>3$. We use this template to calculate {\it
k}-corrections in the UV--near-IR part of the SED.

For the mid-IR-to-submm part of the SED we use the 185-SED template library
constructed by \citet{Swinbank13}, which includes local galaxy templates from
\citet{Chary01}, \citet{Rieke09}, and \citet{Draine07}, as well as the SEDs
from well studied high-redshift starburst galaxies SMMJ2135--0102 ($z=2.32$)
and GN20 ($z=4.05$) from \citet{Ivison10} and \citet{Carilli11}, respectively.
We fit these templates to our observed far-IR stacked photometry of the LBGs
using a $\chi^{2}$ minimisation approach, but allowing the redshift and
normalisation of the templates to vary. To improve the constraints and
accuracy of the SED fitting, we have also extracted stacked photometry for our
LBG sub-samples based on \textit{Herschel} SPIRE maps of the UDS at 250, 350,
and 500\,$\mu$m from the \textit{Herschel} Multi-tiered Extragalactic Survey
(HerMES; \citealt{Oliver12}). The Level 2 data products from the
\textit{Herschel} ESA archive were retrieved, aligned and co-added to produce
maps. For the stacking, we use the raw map from \citet{Swinbank13}, which has
had the mean flux of 1000 random positions subtracted, effectively removing
any systematic contribution from the background or confusion. In addition,
this map includes re-injected flux from any LBG identified in the 24\,$\mu$m
catalogue used to make a deblended version of the map\footnote{The raw
 SPIRE maps, deblended model, catalogues and residual maps used for this analysis are available at
http://astro.dur.ac.uk/$\sim$ams/HSODeblend/UDS/} in \citet{Swinbank13}
(including 142, 15, and 2 in the $z\sim3$, 4, and 5 samples, respectively).
The stacked results are given in Table~1. We note that significant
(i.e.~$>3$-$\sigma$) detections in all three SPIRE bands are made at $z\sim3$
and $z\sim4$, but not at $z\sim5$. To provide a constraint on the Wien side of the spectrum, we have also
stacked the LBGs in 24\,$\mu$m Multiband Imaging Photometer for \textit{Spitzer}
(MIPS; \citealt{Rieke04}) imaging using a 13\,arcsec aperture radius and
apply the standard aperture corrections.  This yields detections of
the $z\sim3$ and 4 LBGs but not for $z\sim5$ LBG sample, and the
results are tabulated in Table~1. Thus we use the stacked fluxes and errors
from Table~1 for the $z\sim3$ and 4 samples in the SED fitting, but convert
the SPIRE and MIPS fluxes of the $z\sim5$ sample to 3-$\sigma$ upper
limits. The best-fitting templates are shown in Fig.~\ref{far-ir-seds}, along with the
range of models which lie within 1-$\sigma$ of the best fit that are used to
calculate uncertainties on our SED-derived parameters.

We use the best-fitting SEDs to compute the basic properties $L_\mathrm{IR}$
and SFR. To calculate the $L_\mathrm{IR}$ values of our average LBG at
$z\sim3$, 4, and 5, we integrate under the best-fitting SEDs between 8 and
1000\,$\mu$m and tabulate the results in Table~2. We then calculate the
corresponding star formation rate (SFR) for our average LBG (see Table~2)
following \citet{Kennicutt98}: ${\rm SFR}\,({\rm M}_{\odot}\,{\rm
yr}^{-1})=1.7\times10^{-10}\,L_{\rm IR}({\rm L_\odot})$. This relation assumes
that the IR luminosity is predominantly powered by star formation (i.e.~a
negligible contribution from an AGN, which is assumed to be a good assumption
in general for LBGs; e.g.~\citealt{Shapley05}; \citealt{Huang07};
\citealt{Rigopoulou10}) and comes from a starburst less than 100\,Myr old,
with a \citet{Salpeter55} initial mass function (IMF). We find that the
best-fitting SED template for the $z\sim3$ sample has a dust temperature of
$T_\mathrm{dust}=37$\,K and the best-fitting SED template for the $z\sim4$ and
5 samples has $T_\mathrm{dust}=38$\,K. For reference, we also fit the
SPIRE+SCUBA-2 photometry using a simple modified blackbody assuming a
dust emissivity index of $\beta=1.5$ (e.g.~\citealt{Blain02}) with the Wien side of the spectrum
modified by a power law of the form $S_{\nu} \propto \nu^{-1.7}$. The derived
dust temperatures from the SED template fitting are lower (by 5--25\,per
cent, with the percentage difference increasing with redshift) compared to the
modified blackbody method, and the $L_\mathrm{IR}$ values from the modified
blackbody fits are a factor of $\approx2$--3 times higher than the SED
templates, demonstrating the additional systematic uncertainty involved in
far-IR SED fitting that should be taken into account.

For each LBG redshift subset, we also perform median stacking\footnote{A median stack minimizes any bias due to a few number of
radio-loud outliers} of the LBG samples in the UDS Jansky Very Large Array (JVLA) 1.4\,GHz data. The median stacked radio fluxes are
S$_\mathrm{1400}=(2.42\pm0.30), (4.20\pm0.65)$, and $(5.06\pm2.25)$\,$\mu$Jy,
for our $z\sim3$, 4, and 5 LBGs, respectively. The radio fluxes were obtained
using the techniques described in Arumugam (in preparation). In brief, the
fluxes were obtained using the {\sc JMFIT} task within AIPS, where the width
of the Gaussian fit to the stack was fixed to be the full width at half
maximum (FWHM) of the corresponding stack at the positions of sources in a
{\it K}-band catalogue. The whole {\it K}-band catalogue was used to provide a
high S/N image from which the Gaussian parameters could be determined. This is
used over the synthesised beam FWHM to account for bandwith smearing effects.
The radio stacks are simply overplotted on the SEDs in
Fig.~\ref{far-ir-seds} as detections (at
$z\sim3$ and 4) or upper limits ($z\sim5$) using a nominal $>3\,\sigma$
threshold.  We note that the detected radio stacks at $z\sim3$ and 4 lie 1--$3\,\sigma$ higher than
the best-fitting far-IR constrained SEDs would predict, hinting
that either the radio emission in the template is not representative of $z\sim3$--4 star-forming
galaxies and/or that there is some AGN
contribution to the radio fluxes of typical LBGs (although, only $\sim$3\,per
cent of LBGs show any AGN activity; \citealt{Steidel02};
\citealt{Laird06}).

Nevertheless, the LBGs appear to be of the LIRG-to-ULIRG class
($L_\mathrm{IR}\sim$10$^{11}$--10$^{12}$\,L$_\odot$), with corresponding SFRs of several
tens to hundreds of M$_\odot$\,yr$^{-1}$. These results are consistent with
previous work within the photometric uncertainties and systematic errors
involved in SED fitting, but our deep stacked fluxes have enabled us to
provide physical constraints on the far-IR SEDs of canonically-selected
``typical'' LBGs for the first time, and are not limited to just the most
massive or UV-brightest subsets (e.g.~\citealt{Lee12}; \citealt{Davies13}).

\subsection{The evolution of the IR luminosity density of LBGs}

We have shown that LBGs contribute a non-negligible fraction of the
submm background. Our
stacking results provide a new suite of constraints for models of luminosity
density and galaxy formation as they directly probe the far-IR density of the
Universe from $z\sim3$ to 5 down to the LIRG level of energy output
($\approx10^{11}$\,L$_{\odot}$).

We can estimate the IR luminosity density of LBGs at $z\sim3$, 4, and 5 by
multiplying the LBG volume densities by the corresponding average LBG
$L_\mathrm{IR}$. To calculate the volume densities, we integrate the UV
luminosity functions from \citet{Reddy09}, \citet{Bouwens07}, and
\citet{McLure09} for $z\sim$3, 4, and 5, respectively, down to an absolute UV
magnitude ($M_{1700}$) of $M^{\ast}_{1700}+1$ at each redshift. Our choice of UV
luminosity limit is motivated by the limitation that $M^{\ast}_{1700}+1$ is
approaching the limiting optical depth of the survey, with the $R$, $I$ and
$z$ band sampling the rest-frame 1700\AA\  emission at $z=3$--$5$. Integrating
the luminosity function to a lower luminosity introduces uncertainty since we
enter a regime where the infrared emission of low-(UV)-luminosity LBGs is not
well measured, and therefore their contribution to the IR luminosity budget is
uncertain. Limiting to $M^{\ast}+1$ ensures a reasonably conservative estimate
of the LBGs' contribution to the luminosity density, whilst sampling a
representative proportion of the population, below the knee in the luminosity
function. Obviously, integrating further down the UV luminosity function
whilst applying the same canonical average $L_{\rm IR}$, would result in a
higher luminosity density, but it is likely that generally $L_{\rm IR}$ is
falling with $L_{\rm UV}$.

\begin{figure} \includegraphics[height=0.45\textwidth,angle=-90]{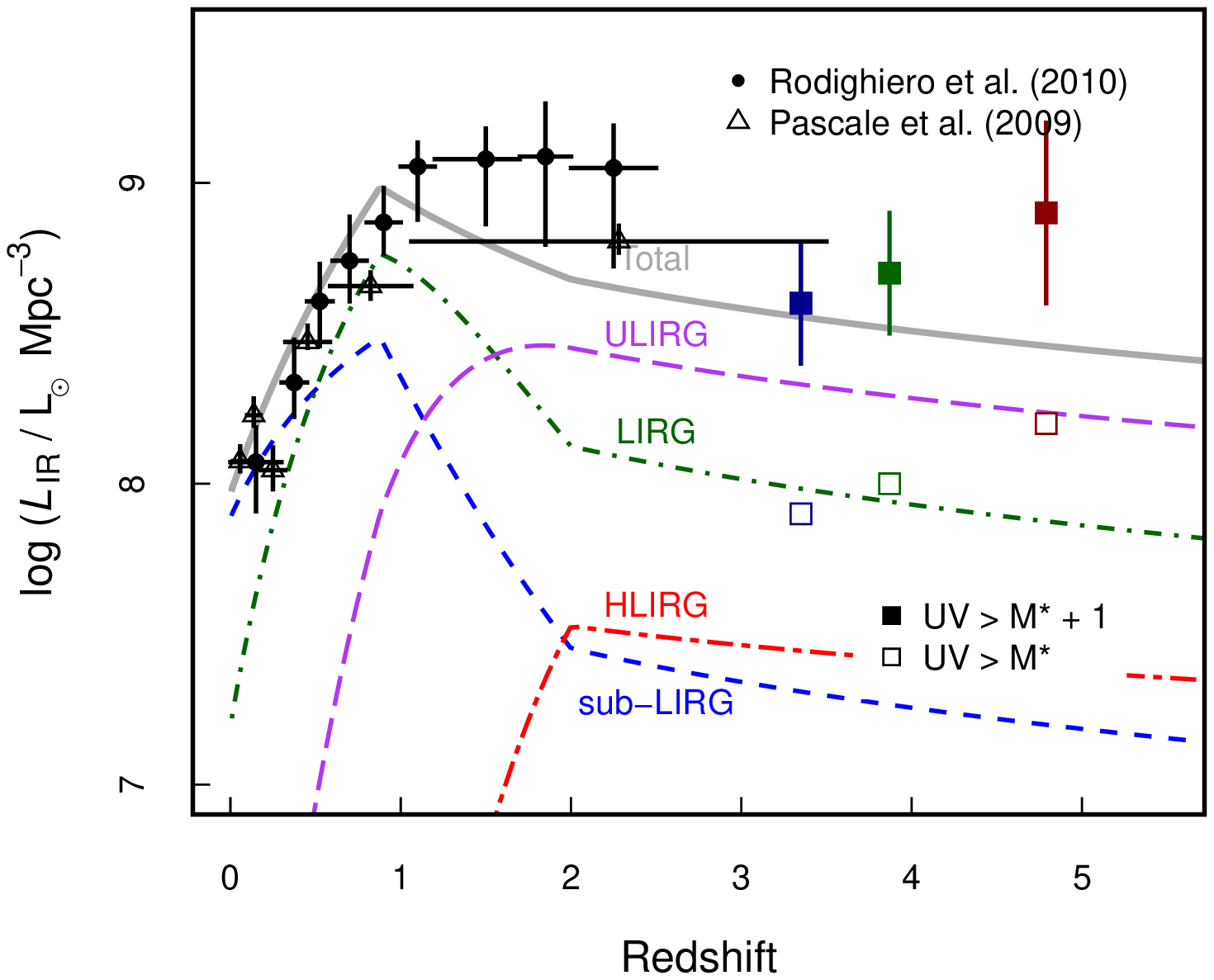}
\caption{Evolution of the
bolometric IR luminosity density (solid line) as a function of redshift from
the parametric backward evolution model of \citet{Bethermin11}. Also shown are
the individual contributions in the models from normal galaxies
($L_\mathrm{IR}<10^{11}$\,L$_{\odot}$), LIRGs
($10^{11}<L_\mathrm{IR}<10^{12}$\,L$_{\odot}$), ULIRGs
($10^{12}<L_\mathrm{IR}<10^{13}$\,L$_{\odot}$), and HyLIRGs
($L_\mathrm{IR}>10^{13}$\,L$_{\odot}$). Measurements from \citet{Pascale09}
and \citet{Rodighiero10} are overplotted, along with our results. To derive
the IR luminosity density, we have multiplied the space density of LBGs at
each redshift by the average $L_{\rm IR}$ we have measured, which is obviously
affected by the choice of luminosity integration limit; we show the effect in integrating down to $M^\ast$ and $M^\ast+1$. These results 
extend observational constraints of the models from $z=3$ to 5, 
showing that the observations of LBGs are broadly consistent with
the model predictions for the LIRG class of galaxies for our $z\sim3$ and 4 LBGs and with the ULIRG class for the $z\sim5$ LBGs.}
\label{bethermin}
\end{figure}

For LBGs with UV luminosities above $M^{\ast}_{1700}+1$, we find corresponding
IR luminosity densities of $(4.0^{+2.3}_{-1.5})\times10^{8}$,
$(5.0^{+2.9}_{-1.8})\times10^{8}$, and
$(7.9^{+7.9}_{-4.0})\times10^{8}$\,L$_{\odot}$\,Mpc$^{-3}$ for $z\sim3$, 4,
and 5, respectively. In Fig.~\ref{bethermin}, we have compared our results
alongside the models of the IR luminosity density from \citet{Bethermin11} and
other available measurements from the literature (\citealt{Pascale09};
\citealt{Rodighiero10}). \citet{Pascale09} stacked 24\,$\mu$m-selected
sources, limiting their study to $z<1.5$--2, in the submm using BLAST data
\citep{Devlin09}. Here, using the efficient canonical LBG selection we are
able to reliably extend the model constraints to $z\sim3$ and beyond for the
first time. It appears that our measured IR luminosity densities for
canonically-selected LBGs at $z\sim3$--5 lie mildly high on average, but are
broadly consistent with the \citet{Bethermin11} model predictions of LIRGs at
$z\sim3$--4 and ULIRGs at $z\sim5$. Obviously the derived luminosity density
depends on our choice of luminosity limit when integrating the UV luminosity
function, since in this calculation we are assuming a fixed average $L_{\rm
IR}$ for all LBGs at each epoch. We have chosen $M^{\ast}_{1700}+1$ to roughly
match the depth of our catalogue, however we show the effect of integrating
down to $M^{\ast}_{1700}$ in Fig.~\ref{bethermin}, which bring the luminosity
densities more in line with the predictions of \citet{Bethermin11} for
galaxies of this luminosity class. Interpreting these observations as a lower
limit to the total infrared luminosity density at $z\approx3$--$5$, and
assuming that the luminosity density peaks at $z\approx1$--$2$, our
observations are consistent with little, or slow, evolution in the infrared
luminosity density over $z\approx3$--$5$.

\subsection{Variation of submm luminosity with stellar mass}\label{stellarmass}

Although the LBG approach is efficient at selecting relatively normal, massive
galaxies at high-redshift, it is important to consider that the `LBG'
population is rather a broad demographic. By definition, the rest-frame
ultraviolet/optical colours of LBGs are quite uniform, as enforced by the
colour selection, but the lack of constraints on the observed near-infrared
(and longer wavelength) properties results, not surprisingly, in a rather
large range of (rest-frame) optical/near-IR colours in LBG samples
(e.g.~\citealt{Rigopoulou06}). As pointed out in \citet{Davies13}, there could
be a substantial variation in the average submm flux density within each of
the LBG samples that we calculated in Section~\ref{stack}. This can be
investigated directly by examining the individual submm flux densities of the LBGs.
Unfortunately, the 1-$\sigma$ scatter in the flux densities of individual sources in each LBG
subset is similar to the map noise ($\approx2.0$\,mJy), indicating that we do
not have sufficient signal-to-noise to detect any real variation or intrinsic
scatter in the submm properties of the bulk of the LBG population.
Nevertheless, a reasonable expectation is that the far-IR luminosity could be
a strong function of stellar mass (e.g.~\citealt{Davies13}), which we now
investigate.

\begin{figure}
\includegraphics[width=0.475\textwidth]{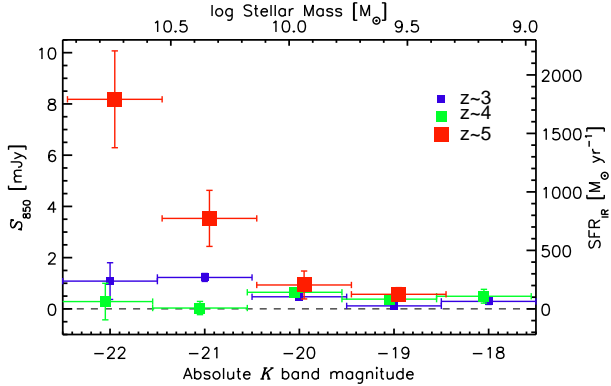}
\caption{Measured 850\,$\mu$m flux density (as a proxy for IR
  luminosity or SFR, shown on right-hand axis, using an average
  approximate conversion from our best-fitting SEDs
  of 1\,mJy$=220$\,M$_\odot$\,yr$^{-1}$) as a function of absolute
  rest-frame $K$-band magnitude (as a proxy
  for stellar mass, shown on the top axis, assuming an average
  mass-to-light ratio here for illustration purposes for the LBGs of 1.15, which assumes a
  \citealt{Salpeter55} IMF).  The plot
  reveals a positive correlation between submm flux density and
  stellar mass at $z\sim5$, with the most massive LBGs being
  more submm luminous on average.}
\label{mass}
\end{figure}

The UV/OIR SEDs described above provide a means of selecting LBGs within each
redshift bin by stellar mass, thus allowing us to apply the appropriate {\it
k}-correction to estimate the absolute rest-frame $K$-band luminosity from the
observed $K$-band magnitude. Using UKIRT filters, the {\it k}-corrections are
2.6, 2.8, and 2.8\,mag for $z=3$, 4, and 5, respectively. Adopting $M_{K}$ as
an empirical proxy for stellar mass (thus allowing us freedom in the
interpretation by assuming different $M_\star/L_{K}$), we repeat the stacking
procedure in bins of $M_{K}$. The results are shown in Fig.~\ref{mass}, which
reveals a positive correlation between submm flux density and stellar mass at
the highest redshifts, with the most massive LBGs at $\sim5$ tending to be the
most submm luminous on average. There is also a mild hint of evolution in the
data -- with the $z\sim5$ LBGs being more IR luminous than similarly massive
LBGs at $z\sim3$ and 4 -- albeit not at a very significant level
($\simeq2\,\sigma$). On closer inspection, we find that the stacked submm flux
result for the highest mass bin in the $z\sim5$ sample is dominated by a
single $\sim4$\,$\sigma$ submm source in the map at one of the LBG locations.
We discuss this further in Section~\ref{discuss}.

\subsection{Comparison of submm-derived SFRs with UV-derived SFRs}

For star-forming galaxies, we expect that the rest-frame UV continuum and the
far-IR emission should be correlated if $\tau\leq1$, since for a given dust covering
fraction, the UV continuum emission should on average scale with the energy
re-radiated in the far-IR. It also follows that the dust absorption should be
correlated with UV reddening, and this has been seen for local starburst
galaxies \citep{Meurer99} and at higher redshifts from $z\sim$1--3
(e.g.~\citealt{Reddy06, Reddy12}; \citealt{Magdis10a,Magdis10b};
\citealt{Heinis13}). It is therefore difficult to distinguish between a galaxy
with a truly high SFR and one with an intrinsically low SFR and large
photometric errors. Correcting for dust attenuation using only optical data is
inherently difficult and uncertain, given the degeneracy between the average
age of the stellar population and affect of reddening, and thus the need for
excellent sampling of the UVOIR spectral range. Submm observations provide a
clean route to assessing the level of dust extinction, and hence the amount of
star formation that is ``obscured'' from view in the optical.

\citet{Shapley01} discuss that the more intrinsically luminous LBGs appear to
be dustier, with redder colours (and should therefore be brighter in the
submm). \citet{Chapman00} reported that submm-bright LBGs tended to have
extremely red colours compared to the average of the population. This was
confirmed by \citet{Reddy12} who showed using \textit{Herschel} 100 and
160\,$\mu$m data that the reddest of their 146 LBGs at $z\sim2$ indeed contain
more dust, finding remarkable agreement between the local and high-redshift UV
attenuation curves (e.g.~\citealt{Meurer99}). With our large LBG catalogues
and deep SCUBA-2 map, we can now explore these issues in more detail and with
greater statistical significance. Here we focus on our largest LBG sample at
$z\sim3$ since this is where we have sufficient S/N to split up the sample
further. To estimate the rest-frame UV (1700\AA) luminosities, we
simply calculate the luminosity from the observed flux in the
observed $R$-band, with no {\it k}-correction.  The associated
UV-derived SFRs are obtained by employing the relation in \citet{Kennicutt98}
to the absolute $M_\mathrm{1700}$ magnitudes of the LBGs (see Table~2). 

\begin{figure}
	\includegraphics[width=\linewidth]{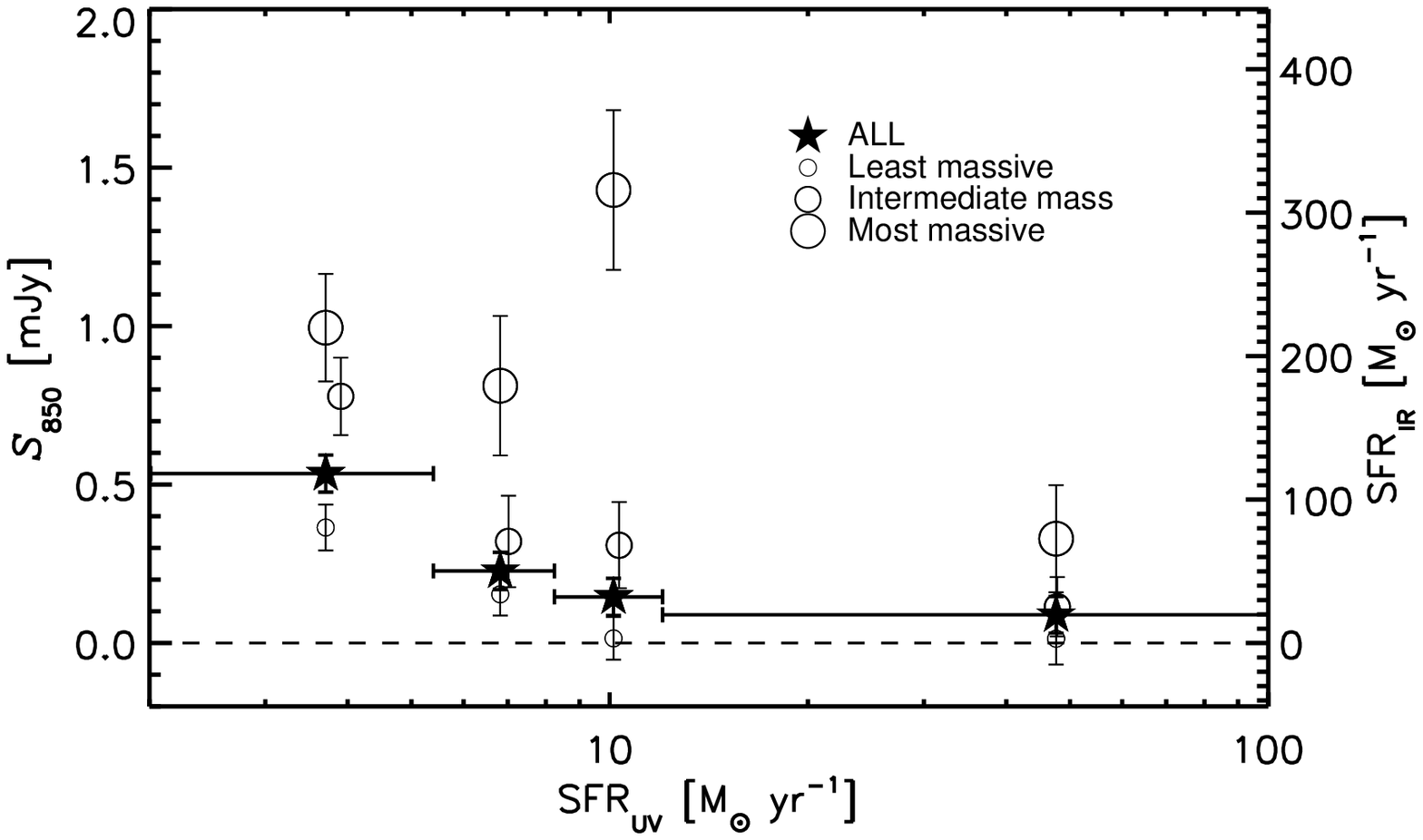}
        \includegraphics[width=\linewidth]{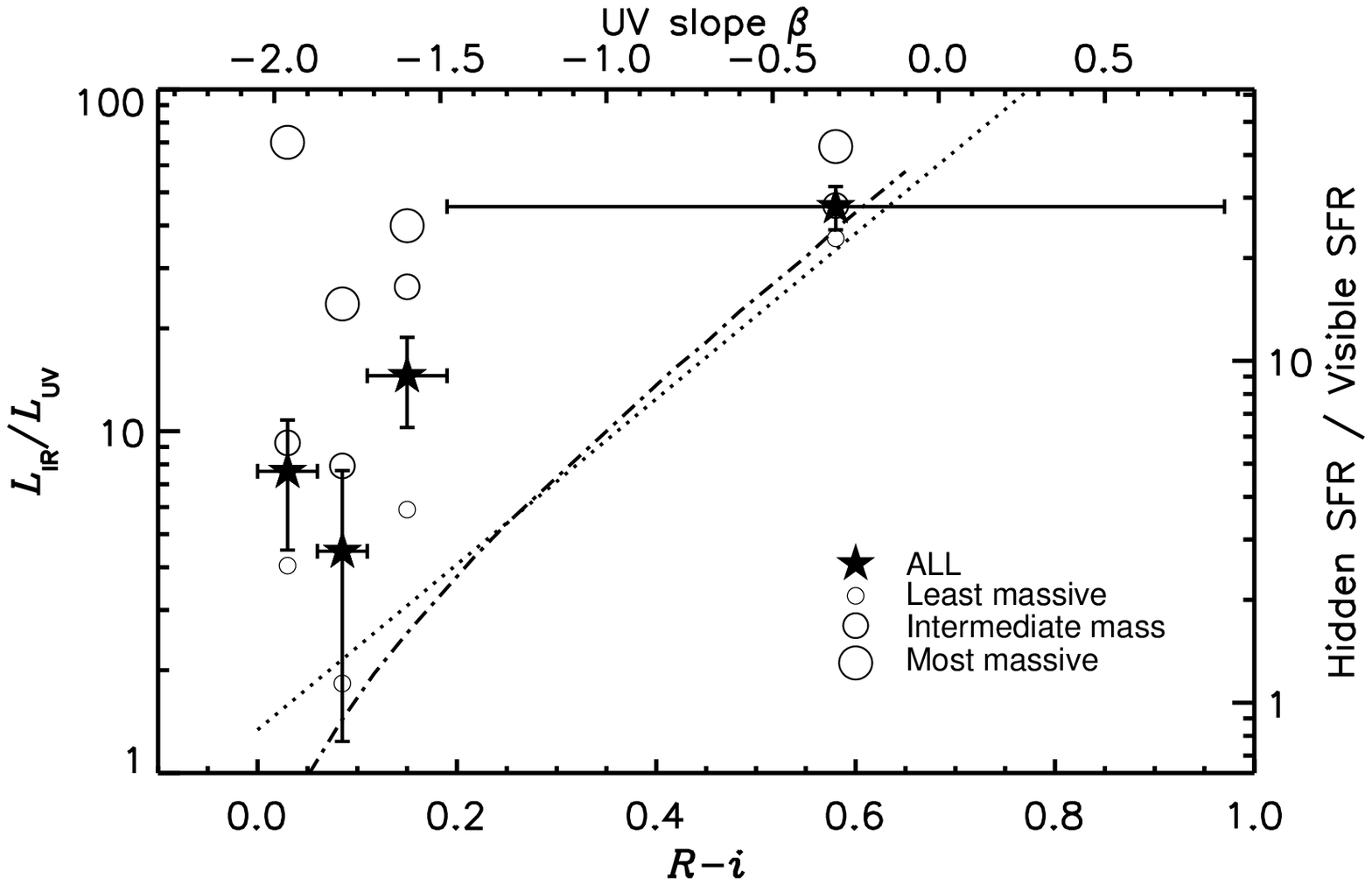}
	\caption{Trends of submm flux, optical colours, and
          UV-estimated SFRs. \textit{Top panel} 
The weighted mean 850\,$\mu$m flux density as a function of
  UV-estimated SFR for LBGs at $z\sim3$, showing an increase
in submm flux density with lower UV SFR.  The error bars represent the standard
error of the mean.  Approximate corresponding IR-SFRs are given on
the right-hand axis for reference and are calculated using an average approximate conversion
  of 1\,mJy=220\,M$_\odot$\,yr$^{-1}$ from our SED fitting results for
  the $z\sim3$ sample.
  \textit{Bottom panel} The mean $L_\mathrm{IR}/L_\mathrm{UV}$ as a function of $R-i$ colour for
different mass bins (with the approximate corresponding UV slope, $\beta$, given on
the top axis for reference), showing that $z\sim3$ LBGs with the reddest colours are
also the most dust obscured, and that this trend scales up with
mass. The error bars represent the standard
error of the mean.  The dotted curve shows the expected $L_\mathrm{IR}/L_\mathrm{UV}$
as a function of optical reddening, assuming an SMC-like extinction
curve (for reference, the dot-dashed curve of \citealt{Meurer99} is also shown).  Taken together, these plots
show directly that for a fixed mass, the reddest LBGs are the most
submm luminous (i.e.~are the most dust obscured), with the average submm
luminosity-to-UV luminosity ratio increasing with mass.}
\label{sfr}
\end{figure}

In Fig.~\ref{sfr}, we plot measured 850\,$\mu$m flux (as a proxy for
obscured SFR) versus the UV-derived SFRs (uncorrected for dust) -- this provides a
measure of the amount of ``dust-obscured'' versus ``unobscured'' or ``visible'' star formation. We find an inverse
correlation of submm flux with UV SFR, that is, the LBGs that are faintest in
the UV are also the most submm bright on average. Are the faint UV population
intrinsically the dustiest or is this simply a mass selection effect? To
explore this, we split up our sample into three bins of equal mass width
across our range of stellar mass: $M_{\star}<5\times10^{9}$\,M$_\odot$ (least
massive); $5\times10^{9}\leq M_{\star} \leq 1\times10^{10}$\,M$_\odot$ (intermediate
mass); and $M_{\star}>1\times10^{10}$\,M$_\odot$ (most massive). At a fixed
mass, we again see that submm flux density increases with decreasing UV SFR,
but now also that the average submm flux increases with the stellar mass. 

What is the distribution of obscured-to-unobscured star formation in the LBGs?
We can explore this by plotting the average $L_\mathrm{IR}/L_\mathrm{UV}$
ratio versus (observed) $R-i$ colour for the LBGs, using the same stellar mass
binning as above (Fig.~\ref{sfr}). This plot demonstrates that on average,
LBGs have $L_\mathrm{IR}/L_\mathrm{UV}\sim5$, but for the reddest
($\beta\approx-0.3$) LBGs in all mass subsets the obscured-to-unobscured ratio
is up to an order of magnitude larger than this, with
$L_\mathrm{IR}/L_\mathrm{UV}\sim50$. As Fig.~\ref{sfr}  shows, for the
populaton as a whole, the increase in
$L_\mathrm{IR}/L_\mathrm{UV}$ is broadly in agreement with that
expected if we assume an SMC-like extinction curve and assume that all
of the optical exinction is caused by dust that is absorbing and
re-emitting the UV radiation field. However, we also see evidence that there is a variation in the
$L_\mathrm{IR}/L_\mathrm{UV}$--$\beta$ trend with mass,
with the most massive LBGs showing the highest $L_\mathrm{IR}/L_\mathrm{UV}$
for all colour bins. The combination of these two plots demonstrates that (a)
at a fixed mass, we can relate optical reddening, as parameterised by the UV
continuum slope $\beta$, to dust-reprocessed emission in the observed submm,
and (b) the most massive LBGs also tend to have higher obscured-to-unobscured
($L_\mathrm{IR}/L_\mathrm{UV}$) ratios (are more extinguised) on average than
the less massive LBGs, hinting at different dust properties/geometries
across the mass range. We discuss this further below.

\section{Discussion: exploring the LIRG population at $\mathrm z>$3--5}\label{discuss}

\begin{figure}
\includegraphics[width=0.475\textwidth]{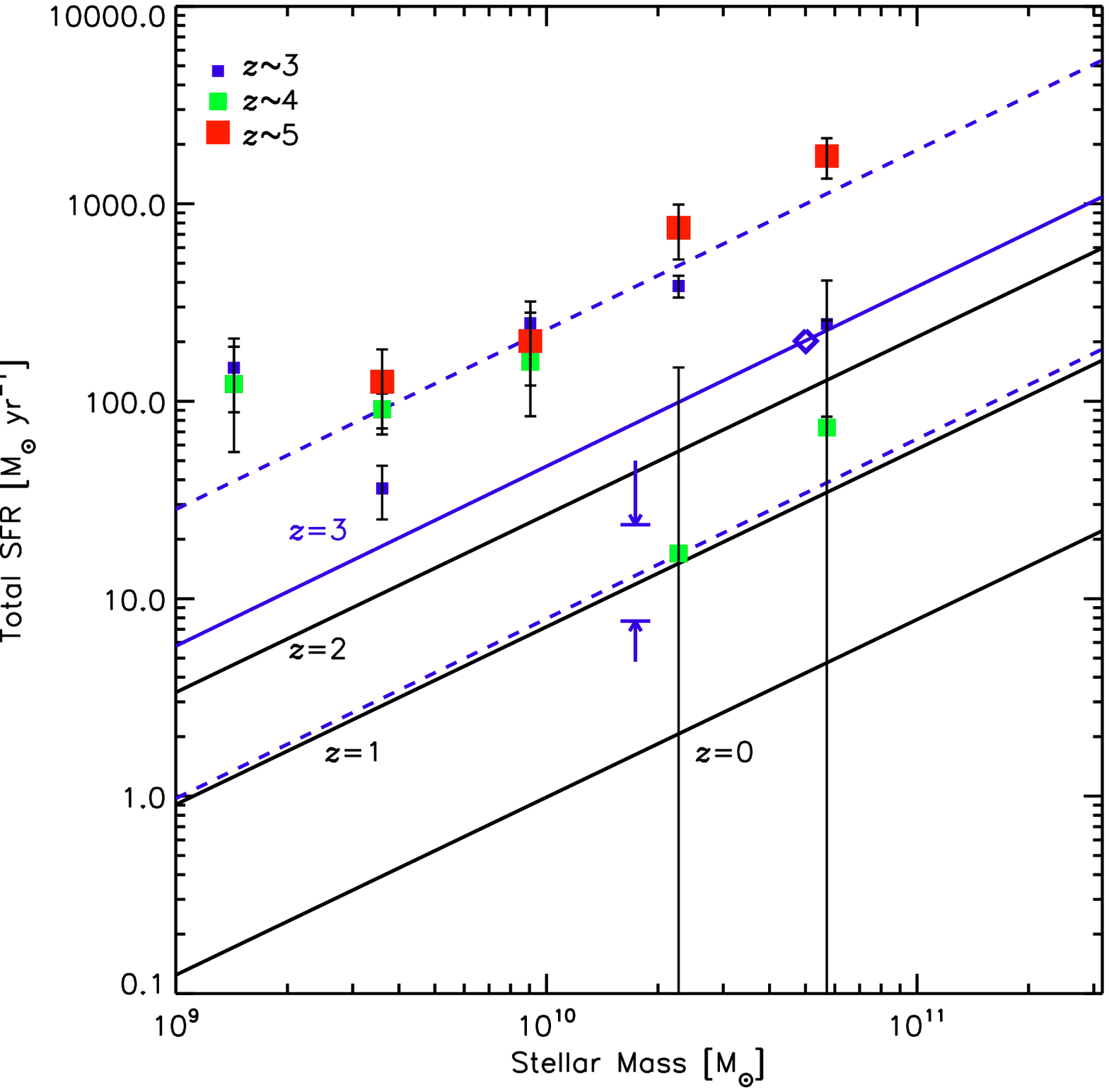}
\caption{The ``main sequence'' of star-forming galaxies predicts
  increasing star formation activity with stellar mass, and the
  relation is known to evolve with redshift.  The observed
  correlations are plotted at $z=0$ \citep{Noeske07}, $z=1$ \citep{Elbaz07}, $z=2$
\citep{Daddi08}, and $z=3$ (from \citealt{Magdis10a}, based on
IRAC-detected LBGs.  Blue lines, with the boundaries shown as
dashed lines, are shown for reference.  This plot has been adapted
from \citet{Davies13} and their limits for typical $z\sim3$ LBGs are
shown as blue arrows, as well as their submm detection of IRAC-22.5
(high mass) sub-sample.  All of the data points
and correlations shown take into account the total
(obscured+unobscured) SFRs.  Similar to Fig.~\ref{mass} (which only
shows the trend for the obscured SFRs with stellar mass), our new data show
that LBGs at $z\sim3$, 4, and 5 indeed demonstrate a ``main sequence''
of star formation activity, with the most massive LBGs tending to have
the highest overall SFRs.  Our data are also consistent with the main
sequence showing no or little further evolution in SFR for similarly
massive LBGs from redshift 3 to 5. }
\label{MS}
\end{figure}

We detect a statistical signal in the submm for the LBG populations at
$z\sim3$, 4, and 5. These stacked fluxes tell us about the ``average'' submm
properties of LBGs at each epoch, and the mean submm flux density tends to increase with
redshift.  A more careful analysis reveals that it appears to be a simple mass-selection
effect -- at higher redshifts lower-mass (and lower overall luminosity)
objects drop out of the sample, mimicking the trend that LBGs increase in IR
luminosity with redshift. When this is taken into account, by splitting our
redshift sub-samples further into stellar mass bins, the data show no
convincing evidence of strong redshift evolution in submm flux (as a proxy for bolometric
luminosity) with absolute rest-frame $K$-band flux (as a proxy for stellar mass).
For comparison to the well-established ``main sequence'' of star-forming
galaxies, which is seen to evolve with redshift,  we plot our
$z\sim3$, 4, and 5 LBG total (unobscured+obscured) SFRs
versus stellar mass in Fig.~\ref{MS}, alongside observed correlations
from $z=0$--3 (\citealt{Noeske07}; \citealt{Elbaz07};
\citealt{Daddi08}; \citealt{Magdis10a}). When plotted in this way, our data show
that LBGs at $z\sim3$, 4, and 5 indeed demonstrate a ``main sequence''
of star formation activity, with the most massive LBGs tending to have
the highest overall SFRs, but again with the main sequence showing no
(or little) further evolution in SFR for similarly massive LBGs from
redshift 3 to 5.

Because we are not prone to dust obscuration effects in the submm (in contrast
to the UV/OIR), we can place strong and definitive constraints on the submm
background for the first time in this luminosity and redshift regime. We have
estimated that LBGs at $3<z<5$ contribute 6--8\,per cent of the 850\,$\mu$m
background, which rises to 14--20\,per cent once we have corrected for our
narrow redshift slices across this range. To put this in perspective, the
bright ($S_{850} >2$\,mJy) SMG\footnote{Here, we consider an SMG to be a
galaxy with $S_{850}\geq$1\,mJy, since for the bulk of these sources,
$S_\mathrm{IR}>S_\mathrm{UV}$, which makes a natural dividing line between
whether the submm or UV is more important energetically.} population makes up
20--30\% (e.g.~\citealt{Coppin06}; \citealt{Weiss09}; \citealt{Swinbank13}) of
850\,$\mu$m background, and the contribution from $>1$\,mJy SMGs peaks at
$\sim50$\,per cent of the SFRD at $z\sim2$ \citep{Wardlow11}. It was thought
that if SMGs and LBGs are essentially the same population, but with LBGs just
being fainter and more numerous, then LBGs could make up the rest of the
``missing'' submm background at 850\,$\mu$m (75--80\,per cent;
\citealt{Adelberger00}). Our measurement shows that at $3<z<5$ LBGs with
(dust-uncorrected) UV luminosities greater than $M^\star_{\rm 1700}+1$
contribute at most 20\,per cent of the 850\,$\mu$m submm background, implying
that the majority of the background is emitted at $z<3$. Of course the
remaining $\sim$50--60\,per cent of the 850\,$\mu$m background, unaccounted
for by SMGs and LBGs, could be feasibly contributed by star-forming galaxies
at lower redshifts, e.g.\ BX/BMs at $z\approx2$, or less UV/optically luminous
``normal'' galaxies (Smith et al.\ in preparation).

Now that we know the average 850\,$\mu$m
flux of the LBG populations at different epochs, we can test this conjecture by looking at the amount of overlap of LBGs and SMGs in the 850\,$\mu$m
number counts. Turning to the deepest and most tightly constrained
850\,$\mu$m number counts available from the cluster lensing fields of \citet{Zemcov10}, at our average detected stacked flux of
$\approx0.25$\,mJy, we find that the surface density of $z=$3 LBGs is
$\approx35$\,per cent that of the detected SMGs.  We find that the LBGs comprise
increasingly smaller fractions of the SMG population as we probe to
higher 850\,$\mu$m fluxes, namely $\approx15$\,per cent at $\approx0.4$\,mJy for the $z\sim4$ LBGs
and $\approx3$\,per cent at $\approx0.9$\,mJy for the $z\sim5$ LBGs.  It
appears that we are finally starting to see a unifying link between the LBG
and SMG populations, with LBGs becoming increasingly important contributors to the submm
background as we probe down into the sub-mJy regime. Note that some classical SMGs are also LBGs, or at least
have UV/optical properties consistent with LBGs
(\citealt{Chapman05}; \citealt{Simpson14}), again indicating overlap
between the populations.
Further progress in unifying the SMG and LBG populations can be made
by examining our candidate SMG-LBG counterparts, which will appear
in a future paper.

We also find that the submm flux increases with LBG optical faintness and
redness -- that is to say, LBGs appear to be dustier and redder on average as
they become more optically faint. This may seem at odds with
\citet{Shapley01}, who claim that more intrinsically luminous LBGs appear to
be dustier, with redder colours. But recall that our UV-SFRs have not been
corrected for dust attenuation, so their intrinsic luminosities will be higher
once the obscured SFRs (as measured in the submm) have been factored in. Submm
flux can be taken as a proxy for the bolometric luminosity of an
LBG\footnote{This is so because of an effect called the negative
$k$-correction, which applies for a galaxy of a given luminosity from
$z\approx$1--8 \citep{Blain02}.}. So in fact, we do see that the most
bolometrically luminous LBGs (as traced in the submm) contain the most dust
(by definition) and have redder optical colours on average, but these are not
the most optically luminous LBGs. Surveys trying to detect LBGs in the submm
individually should therefore focus on the optically faintest and/or reddest
LBGs.

\section{Conclusions}

We have presented an 850\,$\mu$m stacking analysis of LBG samples at $z=3$, 4,
and 5, to measure the average rest-frame far-infrared ($\sim$150--200\,$\mu$m)
luminosity of ``normal'' galaxies in the first 1--2\,Gyr after the Big
Bang. We have several main findings:

\begin{enumerate}

\item The average 850\,$\mu$m flux density of LBGs is $\langle
  S_{850}\rangle\approx$0.2--0.9\,mJy at $z=3$--$5$ (increasing with
  redshift), and they contribute up to 20\,per cent of the submm background.

\item Assuming reasonable templates for the broad (8--1000\,$\mu$m) infrared
spectrum, the results imply that LBGs straddle the LIRG/ULIRG
population, with $L_\mathrm{IR}\sim3$--$11\times10^{11}$\,L$_\odot$
(increasing with redshift).  Our observed evolution of the IR luminosity
density of LBGs is broadly consistent with model predictions.

\item We see a positive correlation between submm flux
density and stellar mass: the most massive LBGs at all three epochs
are the brightest at 850\,$\mu$m.  But we do not see evidence in our
data for a continued strong evolution of the main sequence of star
formation from $z=3$--5.

\item  We have determined that for a fixed mass, the optically reddest
  LBGs are the most submm luminous (i.e.~are the most dust-obscured, by
  definition), and that the average submm luminosity-to-UV luminosity
  ratio increases with mass.

\end{enumerate}

These results provide an empirical baseline to motivate and guide
future direct detection and mapping experiments with the Atacama Large
Millimeter Array (ALMA).  
ALMA will achieve the sensitivities required to detect the dust continuum in
typical individual LBGs in order to fully characterize these sources
which are contributing to the submm background.

\section*{Acknowledgements}

KEKC thanks the Science and Technology Facilities Council of the United
Kingdom (STFC) and JEG thanks the Royal Society for a University Research
Fellowship. JSD acknowledges the support of the European Research
Council via the award of an Advanced Grant,
the support of the Royal Society via a Wolfson Research Merit
Award, and the contribution of the EC FP7 SPACE
project ASTRODEEP (Ref.No: 312725). AK acknowledges support by the
Collaborative Research Council 956, sub-project A1, funded by the
Deutsche Forschungsgemeinschaft (DFG). We would like to thank Alexandra Pope for useful discussions
regarding stacking of clustered populations and Martin Hardcastle for
suggesting to perform the radio stack. This work is based on observations carried out with SCUBA-2 on the
James Clerk Maxwell Telescope (JCMT). The James Clerk Maxwell Telescope is
operated by the Joint Astronomy Centre on behalf of the STFC, the National
Research Council of Canada, and (until 31 March 2013) the Netherlands
Organisation for Scientific Research. Additional funds for the construction of
SCUBA-2 were provided by the Canada Foundation for Innovation.


\end{document}